\documentclass{article}
\usepackage{ pb-diagram, lamsarrow, pb-lams, hyperref}

\def\build#1_#2^#3{\mathrel{\mathop{\kern 0 pt#1}\limits_{#2}^{#3}}}

\usepackage{graphicx}
\usepackage{tabularx}
\usepackage{float}

\begin{document}

\title{The virtual reality framework for engineering objects}

\author{Petr R. Ivankov, Nikolay P. Ivankov}
\renewcommand{\sectionmark}[1]{}
\maketitle
\begin{abstract}
{A framework for virtual reality of engineering objects has been developed.
This framework may simulate different equipment related to virtual reality.
Framework supports 6D dynamics, ordinary differential equations, finite formulas, vector and matrix operations. The framework also supports embedding of 
external software.
}  
\end{abstract}

\section{Introduction}

Problems of virtual reality are indissolubly connected wits other problems of science and engineering. The motion of objects in virtual reality
is depended upon many different engineering related factors. More precise simulation we need, more factors should be taken into consideration.
For instance, currents of artificial satellite equipment interact with magnetic field of the Earth \cite{magneticangularmomentum}, 
so the field and currents should be simulated.
Then we recall that the satellite has a spin-stabilization system \cite{spinstab} and represent it too. If we concern with an aircraft, 
the influence of 
electromagnetic field of the Earth becomes inessential, the motion is mostly depended on aerodynamics and engine's control system behavior. Going
further, we can consider, that rockets, spacecrafts and aircrafts could are deformed, and thus elasticity should also be simulated
\cite{elasticity}

If we suppose the software to be useful for as wide circle of tasks as it possible, it should enable potential
inclusion of simulation from different branches of science and engineering. 
Is it possible? You can download and evaluate interdisciplinary software from
following page 
 
\url{http://www.genetibase.com/universal-engineering-framework-7.php}

This reference also contains examples of applications of this code.

\section{Principles of the framework}

Described framework is based on three main principles. First one is component approach. Second principle is insertion of math formulas. 
Third principle is openness
of framework. So let us consider them.

\subsection{Component approach}

The best method of complicated phenomenon grasping is  decomposition. The best decomposition method is, in 
authors' opinion, a representation of the whole picture by objects and arrows, where latter reproduce interactions 
between objects. It is obvious for the reader aquainted with mathematics that the author of the project has been
inspired by Category Theory \cite{categorytheory}. Furthermore, any object can belong to a set of domains. For example a source of physical field
\cite{field} has a geometric position. Hence it is a subject of positioning domain. This object may be linked to 
other object of positioning domain by positioning links. If a source of a physical field receives and then 
transmits information then it is an information consumer. So the field is also a subject of information domain 
and may be linked to sources of information. And at last it is a subject of physical field domain.
Typical picture of objects (components) of virtual reality simulation is presented on Figure 1.
\begin{figure}[h]
\begin{center}
\hspace{-1cm}
\includegraphics[scale=0.5]{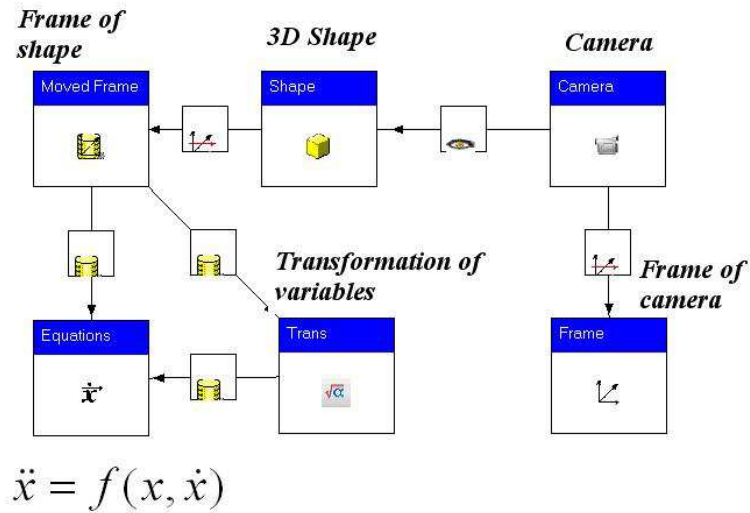}
\caption{Typical example of virtual reality}

\end{center}
\end{figure}

This picture shows motion of 3D shape. We have ordinary differential equations of the shape motion

\begin{equation}
\ddot{x}=f(x, \dot{x}). \nonumber \
\end{equation}

The component at left bottom is a solver of these equations. It is a source of information. In this situation
we should perform some transformation of this information. To do this we use ``Transformation of variables".
Latter component is a consumer of information of the solver. By this means it is connected to solver with 
information link. We also have a ``Frame of shape". It is a moved reference frame that uses information form 
the solver and from the ``Transformation of variables". So it is an information consumer and is connected to its
information providers. The ``3D Shape" is rigidly connected to this frame. The shape is connected to the frame by 
geometrical positioning link. We also have a virtual camera and its reference frame. They are connected with 
geometrical positioning link as well. And at last the camera is connected to the shape by visibility link.

\subsection{Formula editor}

As a software is intends to be interdiscipliniary, it is intended to contain a rich formula editor, that would
enable us to use formulas in different tasks appearing in virtual reality. Signal recognition, transformation of 
3d figures, differential equations solving, definition of figures' size and color etc. etc.- well, there is no need to 
explain where formulas are used and thus where formula editor would be useful. We shall also enable the editor to 
work with variables of different types such as real, integer, boolean vector and so on. In fact, in the 
CategoryTheory project \url{https://sourceforge.net/projects/categorytheory/} formula, the editor operates even 
with Galois fields. The formula editor implemented in related projectsis case sensitive and operates with lots 
of different types. For instance, let $\sin (a)$ be a formula of formula editor. 
What is its meaning? If $a$ is a real variable then result of formula is a real value. However if $a$ is an 
array then $\sin (a)$ is also an array of componentwise calculation of $\sin$. 
If $a$ is a real array and $b$ is a real variable then $a + b$ means an array of sums of components of $a$ with $b$.
Any function of formula editor may be a variable or a result of calculation. 
A function as a result is not a value of function but the function itself. This fact seems unusual for those who 
do not know functional analysis. For example if $a$ is a real variable then $f(a)$ means a result of calculation 
of $f$. If $a$ is a function then $f(a)$ is a composition of functions. 
Formula editor supports matrix and vector operations. 
Examples of usage of vector and matrix operations are presented below:
\begin{equation}
f^taf, \nonumber \
\end{equation}
\begin{equation}
(q^{-1}+h)^{-1}, \nonumber \
\end{equation}
\begin{equation}
a\times b. \nonumber 
\end{equation}
These examples contain transposition of matrixes, products of matrixes, inversion of matrixes and vector 
product of 3D vectors. A very good sample of these operations' applications is Kalman 
filter \cite{Kalmanfilter}. In particular this filter is used in motion
control systems. You can download and evaluate example of this filter from:
\url{http://www.genetibase.com/universal-engineering-framework-6.php}
Recently the formula editor have been enlarged with Dirac delta function \cite{deltafunction}. 
The presence of the delta function 
at the right part of the ordinary differential equation shows that the result function is not continuous. 
Following picture shows presence of delta function in formula editor
\begin{equation}
f(t)\delta (t) \nonumber \
\end{equation}

\subsection{Openness of the software}
Usually every developer or company has its own projects of those object domain. 
This software does not require discarding of existing projects. Any object domain project may be included to 
this software. If you wish to include your project or its part, then you should develop an adapter, compile 
the class library, and link it to this software. 
The adapter should contain one or more classes that implement one or more interfaces of this software. 
A more profound description of these interfaces is contained in the developer's guide. You can download the guide from
AstroFrame homepage \url{https://sourceforge.net/projects/astrohalaxy}.

\section{Examples}

This section is a brief review. Profound description could be found in guides. You can  also download and 
evaluate examples from pages devoted to this software.
\subsection{Reference frames}
Reference frame is one of basic notions of virtual reality. We should link visible objects and virtual cameras to reference frames.
Architecture of the framework uses relative frames, those have forest structure. Example of such structure is presented on the
following diagram:

\[
\begin{diagram}
	\node{1.1.1} \arrow{se,t}{}  \node[2]{1.1.2 }\arrow{sw,t}{} \node[2]{1.2.1 }\arrow{sw,t}{} \node[3]{2.2.2 }\arrow{sw,t}{} \\
	\node[2]{ 1.1} \arrow{se,t}{}  \node[2]{1.2 }\arrow{sw,t}{} \node{ 2.1} \arrow{se, t} \node[2]{2.2} \arrow{sw, t}\\
	\node[3]{1} 	\node[3]{2}
\end{diagram}
\]

There exists a zero reference frame. All root nodes of the forest correspond to this frame. 6D positions of other
frames (nodes) are relative to 6D positions of their parents. It means that 6D position of frame 1.1.2 is relative to 6D position
of frame 1.1 etc. This architecture enables us to create different useful situations easily. For example we can install a set of virtual cameras
on air(space)craft. It is easy to visualize a plane with Swing-wing \cite{Swingwing}

\subsection{Motion of objects}

The framework enables us to define motion by finite formulas and ordinary differential equations.
Besides this facilities it contains a special component for simulation 6D dynamics of rigid body.
This component operates with forces, momentums and moments of inertia \cite{Momentofinertia}.

\subsection{Deformation of figures and views}
This framework performs data processing. In particular it is used for deformations of 3D shapes.
Figures 2 and 3 show deformation of a plane.

\begin{figure}[h]
\begin{center}
\hspace{-1cm}
\includegraphics[scale=0.25]{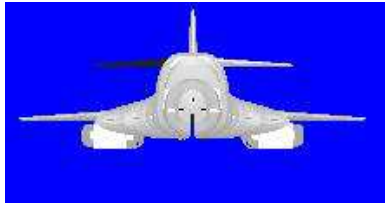}
 \caption {Plane in normal state}
 \end{center}
\end{figure}

\begin{figure}[h]
\begin{center}
\hspace{-1cm}
\includegraphics[scale=0.25]{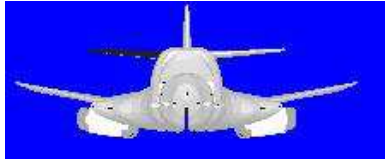}
 \caption {Deformed plane}
 \end{center}
\end{figure}

Besides deformation of 3D shapes we need deformations of views to simulate a view through curved mineral glass, or view in distorting mirror.
\begin{figure}[h]
\begin{center}
\hspace{-1cm}
\includegraphics[scale=0.15]{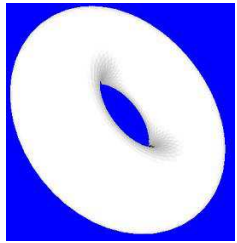}
 \caption {Torus. Normal view}
 \end{center}
\end{figure}

\begin{figure}[h]
\begin{center}
\hspace{-1cm}
\includegraphics[scale=0.15]{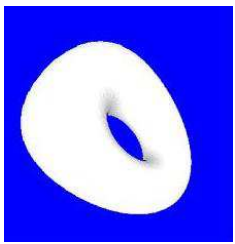}
 \caption {Torus. Deformed view}
 \end{center}
\end{figure}

Figures 4 and 5 shows torus and its deformation of view. Following rules of 2D deformation was used:
\begin{equation}
x\mapsto x^2; \nonumber \
y\mapsto y^2. \nonumber \
\end{equation}

\subsection{Visualization of interaction with fields}

Virtual reality does not operates with realistic pictures only. For example we should visualize invisible physical fields,
temperatures or tenses of materials. Since this framework is interdiciplinary one it enables us to perform this visualization.
\begin{figure}[h]
\begin{center}
\hspace{-1cm}
\includegraphics[scale=0.15]{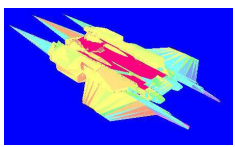}
\caption {Visualization of surface currents}
\end{center}
\end{figure}

Figure 6 presents an example of visualization of surface currents. 

The framework has excellent data processing.
It may be used for simulation of physical fields those have different nature.
The framework also uses data processing for simulation of interaction between fields and surfaces of bodies.
High level of abstraction makes the framework very
flexible. So it enables us to simulate any physical properties of surfaces.
\subsection{Celestial bodies}
Celestial bodies have a set of different applications to virtual reality.
First application is visualization of them. We often should to visulize celestial bodies with spacecrafts and aircrafts.
Second application is usage of physical fields. Gravity fields of Earth, Moon and Sun are taking an influence on the 
motion of artificial satellites. Moreover Earth's magnetic field affects angular motion of satellites. 
Third application is simulation of astronomical instruments \cite{Astronomicalinstruments}
Lots of them are used in control systems of spacecrafts' and aircrafts' motion. Some problems of virtual reality require processing with a lot
of stars. This framework operates with catalogues of stars those are stored in databases. Present version of the framework supports 
ODBC, SQL Server and Oracle drivers for working with databases. Now there exists a lot of different star catalogues.
They have different structure. The framework is compatible with all of them since it can perform any SQL query to database.
Besides SQL query framework have a facility of advanced filtration of stars. This filtration uses data processing that could be
particularly performed with formula editor. Usually star catalogues contain information that cannot be directly used for visualization.
They may contain parallax \cite{Parallax} declination $(\delta)$ and right ascension $(\alpha)$  
\cite{Equatorialcoordinatesystem},
mean BT and/or VT magnitude \cite{NOMAD}. The data processing is used for transformation of this parameters into
stars' coordinates and those visible sizes and color. It is worth to note that there exists such tasks where the simulated
color of the star does not correspond to color that is perceive by human eye. Simulated color could represent astronomical
instruments and may show invisible spectrum of star. Since visible colors are linked to data processing we can simulate any
dependence of represented color on stars' parameters.

\end{document}